\documentclass[a4paper]{article}
\usepackage{multirow}
\usepackage{INTERSPEECH2021}

\title{SpeechMoE: Scaling to Large Acoustic Models with Dynamic Routing \\
Mixture of Experts}
\name{Zhao You$^{*1}$, Shulin Feng$^{*1}$, Dan Su$^1$, Dong Yu$^2$}
\address{$^1$Tencent AI Lab, Shenzhen, China\\
$^2$Tencent AI Lab, Bellevue, WA, USA }
\email{\{dennisyou, shulinfeng, dansu, dyu\}@tencent.com}
\begin{document}

\maketitle

\begin{abstract}
Recently, Mixture of Experts (MoE) based Transformer has shown promising results in many domains. This is largely due to the following advantages of this architecture: firstly, MoE based Transformer can increase model capacity without computational cost increasing both at training and inference time. Besides, MoE based Transformer is a dynamic network which can adapt to the varying complexity of input instances in real-world applications. In this work, we explore the MoE based model for speech recognition, named SpeechMoE. 
%First, to further control sparsity of router activation and improve the diversity of gate values, we propose a sparsity $L1$ loss and mean importance loss respectively. Then, we propose a new router architecture which can simultaneously utilizes the information of a shared embedding network and the hierarchical representation of different MoE layers. Finally, we show that speechMoE can achieve lower word error rate with comparable computation cost than traditional static networks. Specially, our proposed architecture provides up to 10\% relative character error rate improvement over baseline model.
To further control the sparsity of router activation and improve the diversity of gate values, we propose a sparsity $L1$ loss and a mean importance loss respectively. In addition, a new router architecture is used in SpeechMoE which can simultaneously utilize the information from a shared embedding network and the hierarchical representation of different MoE layers. Experimental results show that SpeechMoE can achieve lower character error rate (CER) with comparable computation cost than traditional static networks, providing 7.0\%$ \sim$ 23.0\% relative CER improvements on four evaluation datasets.

\end{abstract}

\renewcommand{\thefootnote}{\fnsymbol{footnote}}
\footnotetext[1]{Equal contribution.}
%$\footnote{Equal contribution.}$

\section{Introduction}
Owing to powerful representation, Deep Neural Networks (DNN) have gained great success in speech recognition \cite{DE2, DE3}.  Various types of neural network architectures have been employed in ASR systems, such as convolutional neural networks (CNNs)~\cite{sainath2013deep, CNN}, long short-term memory (LSTM)~\cite{graves2013hybrid}, gated recurrent unit\cite{ravanelli2018light}, time-delayed neural network \cite{peddinti2015time},  feedforward sequential memory networks (FSMN) \cite{zhang2018deep}, etc. Recently, more powerful deep models such as Transformer\cite{transformer}, Emformer\cite{emformer} and Conformer\cite{conformer} have proved their efficacy to further improve the speech recognition performance. 

Increasing model and training data size has been shown an effective way to improve the system performance, which is especially demonstrated in the field of language modeling~\cite{shoeybi2019megatron, brown2020language}. Recently, deep mixture of experts (MoE) based approaches \cite{jacobs1991adaptive, jordan1994hierarchical} have been intensively investigated and applied in different tasks such as language modeling \cite{lepikhin2020gshard, fedus2021switch} and image classification\cite{gross2017hard, ahmed2016network, wang2020deep, cai2021dynamic}. The benefits mainly come from two aspects: First, MoE is an effective way to increase model capacity. Second, with introduction of the sparsely-gated mixture-of-experts layer \cite{shazeer2017outrageously}, an attractive property of MoE models is the sparsely dynamic routing, which enables us to satisfy training and inference efficiency by having a sub-network activated on a per-example basis.

In real-world applications, speech recognition systems need to be robust with different input conditions such as speakers, recording channels and acoustic environments. Larger models are appealing while the increase of training and inference cost can not be afforded. The major problem is that the computation cost of a static model is fixed and can not be adaptive to the varying complexity of input instances. Therefore, developing mixture of expert models for speech recognition with dynamic routing mechanism is a promising exploration.

In this study, we explore mixture of experts approach for speech recognition. We propose a novel dynamic routing mixture of experts architecture, similar to \cite{fedus2021switch}, which comprises of a set of experts and a router network. The router takes output of the previous layer as input and routes it to the best determined expert network. We find that the balance loss proposed in \cite{fedus2021switch} achieves balanced routing but the sparsity of router activation can not always be guaranteed. Here, we propose a sparsity $L1$ loss to encourage the router activation to be sparse for each example. Besides, we use a mean importance loss to further improve the balance of expert utilization. Furthermore, a shared embedding network is used in our architecture to improve the route decisions, whose output will be combined with the output of previous layers as the input of routers.

%we propose a shared embedding network for improving experts selection decisions.  This will combine the output of previous layers and the output of the shared embedding network as the input of routers.

The rest of the paper is organized as follows. Section 2 reviews the related works of MoE and Section 3 represents our proposed method SpeechMoE. The experimental setup is described in Section 4 and the experimental results are reported in Section 5. Finally, we conclude this paper in Section 6.

%\section{SpeechMoE}
\section{Related works}
In this section, we mainly describe two different architectures of MoE.
%Then, we introduce the detailed model architecture design of our proposed model. 
%In section 2.2, we first introduce the balance loss which is used in switch transformer. Then, we propose two auxiliary loss to further optimize the training process. Finally, the total model loss is described. 
%In this section, we describe the detailed model architecture design and then introduce the loss function formulation.

\label{sec:typestyle}
\begin{figure*}[!tb]
\begin{minipage}[b]{1.0\linewidth}
  \centering
  \centerline{\includegraphics[width=14.5cm]{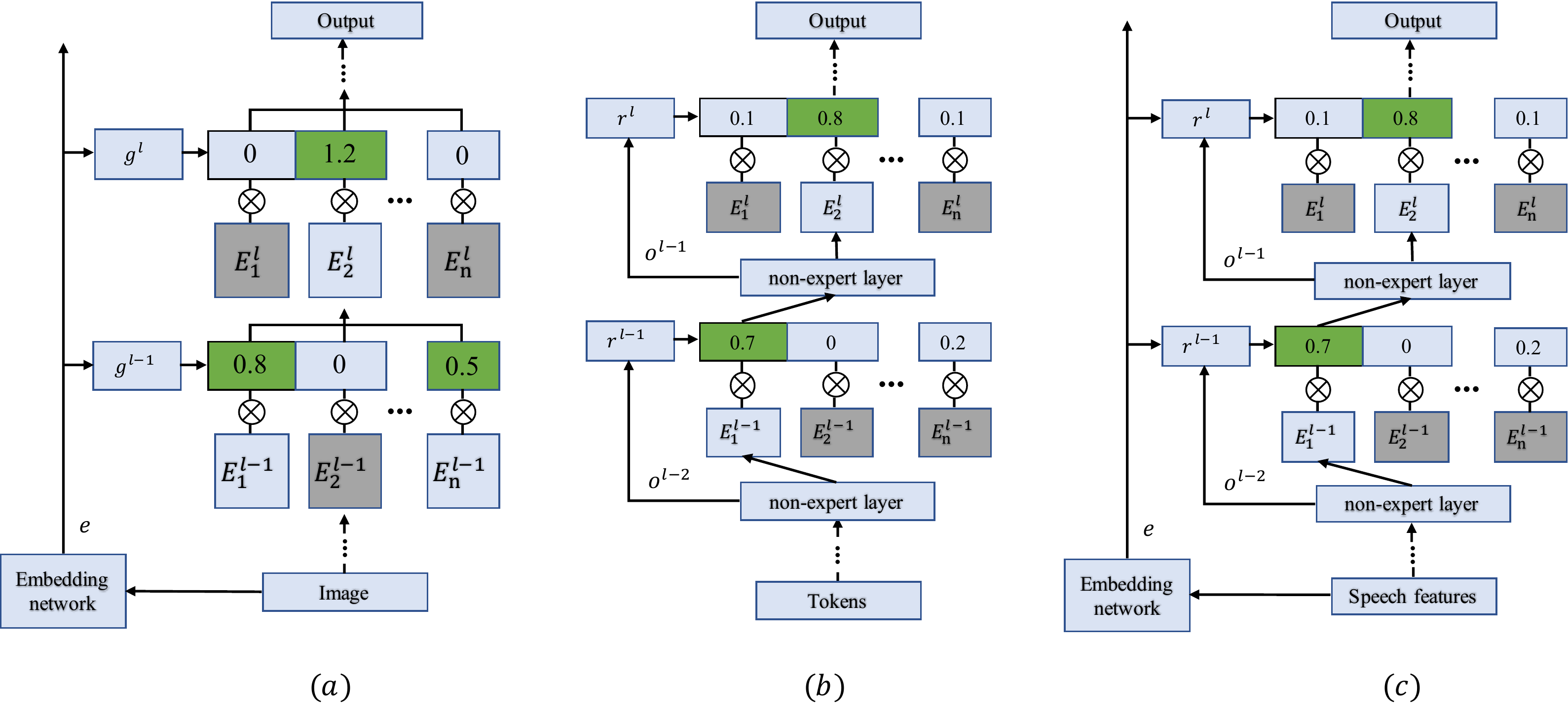}}
%  \vspace{2.0cm}
%  \centerline{(a) Result 1}\medskip
\end{minipage}
\vspace{-15pt}
%\caption{ (a) presents the architecture of the DeepMoE. In DeepMoE, the output of the shared embedding network is used as the input of the gating network.(b) represents the switch transformer architecture where that the output of previous layer is used as the input of the router.(c) presents the architecture of the proposed SpeechMoE. In SpeechMoE, a new router architecture is proposed which concatenates the output of shared embedding with output of previous layer as the input of the router.}
\caption{
(a), (b) and (c) represent the architecture of DeepMoE, Switch Transformer and SpeechMoE respectively. Similar to Switch Transformer, only one expert with the largest router probability in each MoE layer is used in the SpeechMoE, which is different from DeepMoE. Besides, the SpeechMoE utilizes a shared embedding and output of the previous layer as the input of each router. 
}
\vspace{-15pt}
%\caption{Frames accuracy on development set against number of epochs for SGD, HF and SHF. ($N$T means SHF or HF is initialized with the result of SGD after $N$ epochs.)}
%\label{fig:res}
\end{figure*}

%%%\subsection{Model architecture}
%\subsection{Prior Work of MoE}
%\subsection{Related works}
\subsection{DeepMoE}
The DeepMoE architecture proposed in \cite{wang2020deep} can achieve lower computation cost and higher prediction accuracy than standard convolutional networks.  The architecture designs a sparse gating network which can dynamically select and re-weight the channels in each layer of the base convolutional network. Fig.1(a) shows the detailed architecture of DeepMoE. 
The DeepMoE consists of a base convolutinal network, a shared embedding network and a multi-headed sparse gating network. The gating network transforms the output of the shared embedding network into sparse mixture weights:
\begin{equation}
g^{l}(e) = f({W^{l}_{g}} \cdot e)
\end{equation}
where $g^{l}(e)$ is the sparse mixture weights of $l$-th convolutional layer, e is the output of the shared embedding network, and $f$ is the activation operation(i.e., Relu). Then, the output of $l$-th convolutional layer can be formulated as:
\begin{equation}
y^{l} = \sum_{i=1}^{n} g^{l}_{i} E^{l}_{i}
\end{equation}
where $n$ is the input channels number of $l$-th convolutional layer and $E^{l}_{i}$ is the $i$-th channel of $l$-th convolutional layer, treated as the $i$-th expert in $l$-th layer.

The loss function for training DeepMoE is defined as:
\begin{equation}
L(x;y) = L_{b}(x;y) + \alpha  L_{g}(x;y) + \beta  L_{e}(x;y)
\end{equation}
where $x$ and $y$ are the input image feature and target label, respectively. $L_{b}$ is the classification loss, $L_{g}$ is the L1 regularization term which controls sparsity of the gating network and $L_{e}$ is the additional classification loss which encourages the diversity of shared embedding network. 

%Different with switch transformer, the input of gating networks is a high level representation which is learned by a shared embedding network. Thus, for a training sample, the input of gating network of each MoE layer is same. This lead to that the hierarchical representation of different MoE layers can not be used for the gating selection.

\subsection{Switch Transformer}
Fedus et al. proposed the Switch Transformer \cite{fedus2021switch}  for language modeling, which further reduces computation and communication costs by simplifying the MoE routing algorithm. The architecture of Switch Transformer is described in Fig.1(b), where experts refer to feed-forward networks and the non-expert layers refer to the self-attention layers. Each MoE layer consists of $n$ experts and a router layer. It takes output for the previous layer as input and routes it to the top-1 expert with the largest router probability. Let $W^{l}_{r}$ and $o^{l-1}$ be the router weights of the $l$-th layer an the output of the previous layer, then the router probability can be defined as follows:

\begin{equation}
  r^{l} = W^{l}_{r} \cdot o^{l-1}
\end{equation}

\begin{equation}
  {p^{l}_{i}} = \frac{exp^{r^{l}_{i}}}{\sum_{j=1}^{n}exp^{r^{l}_{j}}}
\end{equation}
Then, the selected expert's output is also gated by router probability to get output of the MoE layer,
\begin{equation}
y^{l} = p^{l}_{i} E^{l}_{i}
\end{equation}

Since only one expert is active in each layer, the Switch Transformer can keep the computation cost constant while scaling to a very large model. To encourage a balance load across experts, the balancing loss \cite{fedus2021switch} is added into the loss function and defined as:
\begin{equation}
L_{b}=n \cdot \sum_{i=1}^{n} s_{i} \cdot P_{i}
\end{equation}
where $s_{i}$ is the fraction of samples dispatched to expert i, $P_{i}$ is the fraction of router probability allocated for expert $i$.

\section{SpeechMoE}
\subsection{Model architecture}
Fig.1(c) shows an overview of the architecture of our proposed SpeechMoE. For speech recognition, its input is speech features (e.g. fbanks) and the input frames will be dispatched to experts in each layer. Similar to the Switch Transformer, SpeechMoE only selects one expert in each layer to reduce the computation cost. Compared with Switch Transformer and DeepMoE, the SpeechMoE concatenates the shared embedding with output of the previous layer as the input of routers, which can be defined as:

\begin{equation}
r^{l} = W_{r}^{l} \cdot Concat(e;o^{l-1})
\end{equation}

This router mechanism comes from two considerations: (1) All gating values in DeepMoE are controlled by the shared embedding, which may decay to similar gating results in each layer. Utilizing the hierarchical representation from output of each layer may lead to diverse routing results for SpeechMoE. (2) The shared embedding relative to the goal task may be helpful to get a better routing strategy, providing a high-level distinctive representation and making the experts specialized to process distinct input frames.

\subsection{Training objective}
\subsubsection{sparsity $L1$ loss }
In our study, we find that the router probability distribution tends to be uniform when we only use the balancing loss proposed in \cite{fedus2021switch}, resulting in a bad performance. In order to encourage the sparsity of router activation, we propose a sparsity $L1$ loss, defined as follows:

\begin{equation}
L_{s}= \frac{1}{m}\sum_{i=1}^{m} \parallel{\hat{f}_{i}}\parallel_{1}
\end{equation}
where $\hat{f}_{i}= \frac {f_{i}} {{\parallel} {{f}_{i}} \parallel_{2}}$, stands for the unit normalized router probability distribution of sample $i$, and $m$ is the number of samples in this mini-batch. Due to the unit normalization, minimizing the $L1$ norm will force the distribution close to space axes and attain sparsity.
%where $\hat{f}_{i}= \frac {f_{i}} {{\parallel} {{f}_{i}} \parallel_{2}}$, The sparseness $L1$ loss push the router probability distribution to have the sparsity of activation for each example.

%In this paper, we consider two training loss:CE(Cross Entropy)  and CTC (Connectionist Temporal Classification).  
%It[] has shown that the experts tend to receive the examples from the same class.  We refer it as the same label in a batch of frame level force alignment.  The CE loss tend to generate a uniform distribution for the different labels. Normally, We can achieves uniform routing of different labels. However, the CTC loss tend to generate the shape spike distribution that only
%a few spikes for each output non-blank label while predicting blank label with high probability the rest of time.  

%As for this, we propose an two head router.

%Thereby, the regularized CE loss will help to produce the accurate alignment for the output target while won’t effect the distribution of blank label. 

\subsubsection{Mean importance loss}
We have also observed that model isn't balanced enough when increasing the number of experts. To solve this problem, we use a modified importance loss\cite{shazeer2017outrageously} to replace the balancing loss, defined as follows:
\begin{equation}
Imp = \frac{1}{m}\sum_{i=1}^{m} p_{i}
\end{equation}
\begin{equation}
\bar{L}_{m}= n\sum_{j=1}^{n}  {{Imp}_{j}}^{2}
\end{equation}
The mean importance is defined as the mean activation of experts on batch of samples and the loss is defined as the squared sum of mean importance of each expert. It's clear that when mean importance of each expert is averaged $\frac{1}{n}$, the loss reaches the minimum. Compared with the balancing loss in which $s_i$ is not differentiable, the mean importance loss is more smooth, leading to a more balanced routing strategy.

%We have observed that model converges slower when we increasing the number of experts from 2 to 4. We suspect the reason is that  $s_{i}$ in balance loss which defined in Equation 7 is not differentiable. To solve this problem, we propose a modified importance loss \cite{shazeer2017outrageously} $\bar{L}$. First, we instead use the batch mean of the router values of an expert to define the importance of an expert relative to a batch of training samples. Then, the additional loss $\bar{L}$ is computed with the square sum of the set of importance values. Finally, we use a hyper-parameter $\beta$ as the multiplicative coefficient for the addition loss.

%\begin{equation}
%Imp(x)= \frac{1}{T}\sum_{t=1}^{T} p_{t}(x)
%\end{equation}
%\begin{equation}
%\bar{L}_{m}(x)= \beta \sum_{j=1}^{N}  {{Imp(x)}_{j}}^{2}
%\end{equation}
%From Equation 11, We can observe that $Imp(x)_{j}$ is differentiable and make the model converges normally.

\subsubsection{Loss function}
Given the input $x$ and the target $y$, the full loss function of our method is defined as 
\begin{small}
\begin{equation}
\begin{split}
L(x;y)=L_{r}(x;y) +\alpha L_{s}(x) + \beta \bar{L}_{m}(x) + \gamma  L_{e}(x;y)
\end{split}
\end{equation} 
\end{small}Among these items, $L_{r}$ is the CTC loss\cite{graves2006connectionist} for speech recognition, $L_{s}$ and $\bar{L}_{m}$ are the mentioned sparsity $L1$ loss and mean importance loss, used to encourage sparsity and diversity of the SpeechMoE model. Similar to \cite{wang2020deep}, we introduce an additional embedding loss $L_{e}$, which is also the CTC loss. It shares the same goal with our SpeechMoE model and provides reliable embeddings for the routers. $\alpha$, $\beta$, and $\gamma$ are the scale for $L_{s}$, $\bar{L}_{m}$ and $L_{e}$ respectively.

%where $L_{r}$ is the CTC loss\cite{graves2006connectionist} for speech recognition, which encourages a high sequence prediction accuracy . %$L_{b}$  is the load balance loss which is defined by Equation 4.  $L_{s}$  is the sparseness loss which is defined by Equation 5.  $L_{i}$  is the important loss which is defined by Equation 6.
%$L_{b}$, $L_{s}$ and $L_{m}$ encourage to learn a router that selects a diverse, low-cost mixture of experts for each training sample.
%Finally, we introduce an additional embedding loss $L_{e}$, which is the CTC loss.  
%This encourages the similar embedding have similar router decisions. 
%For all the experiments, we set $\alpha = 0.1$, $\beta = 0.1$ and $\gamma = 0.01$.
%In a word, we must learn

%Further, we use an addtional embedding loss $loss_{e}$, which is the CE loss or CTC loss. Intuitively, this encourage the examples from the frame level class should have similar embedding and thus similar subsequent router decisions. 

\section{Experimental Setup}
\label{sec:format}

\subsection{Training setup}

 The speech features used in all the experiments are 40-dimensional log-Mel filterbank features appended with the first-order and the second-order derivatives. Log-mel filterbank features are computed with a 25ms window and shifted every 10ms. We stack 8 consecutive frames and subsample the input frames with 3. A global mean and variance normalization is applied for each frame. All the experiments are based on the CTC learning framework. We use the CI-syllable-based acoustic modeling method \cite{syllable} for CTC learning. The target labels of CTC learning are defined to include 1394 Mandarin syllables, 39 English phones, and a blank. Character error rate results are measured on the test sets and the floating point operations (FLOPs) for a one-second example is used to evaluate the inference computation cost. We use a pruned, first pass, 5-gram language model. All the systems use a vocabulary that consists of millions of words. Decoding is performed with a beam search algorithm by using the weighted finite-state transducers (WFSTs).

% We observed that training CTC from scratch is unstable and sometimes  training will fail to converge. In order to solve the CTC unstable learning problem, all acoustic models were initialized from CE loss pre-trained models. To get the alignment of CE learning, we first map the existed 12485 CD-Phones to 1433 target labels (1394 Mandarin syllables and 39 English phones).    

\subsection{Datasets}

Our training corpus is mixed data sets collected from several different application domains, all in Mandarin.   In order to improve system robustness,  a set of simulated room impulse responses (RIRs) are created with different rectangular room sizes, speaker positions, and microphone positions, as proposed in \cite{far2}. Totally, It comes to a 10k hours training corpus. 

To evaluate the performance of our proposed method, we report performance on 3 types of test sets which consist of hand-transcribed anonymized utterances extracted from reading speech (1001 utterances), conversation speech (1665 utterances) and spontaneous speech (2952 utterances).  We refer them as Read, Chat, and Spon respectively. In addition, to provide a  public benchmark, we also use AISHELL-2 development set (2500 utterances) recorded by high fidelity microphone as the test set.

\subsection{Acoustic Model}
Our acoustic models consist of four components: MoE layer, sequential memory layer \cite{fsmn}, self-attention layer \cite{you2020dfsmn} and the output softmax layer.
%For the  router layer, the 
Each MoE layer includes a router and a set of experts which is a feed forward network with one hidden layer of size 1024 activated by ReLU and an projection layer of size 512. For the sequential memory layer, the look-back order and look-ahead order of each memory block is 5 and 1 respectively, and the strides are 2 and 1 respectively. For the self-attention layer, we set the model dimension $d = 512$ and the number of heads $h = 8$. For every layer excluding the output softmax layer, the residual connection is applied.

%\subsubsection{Our proposed model}
The backbone of our model consists of 30 MoE layers, 30 sequential memory layers and 3 self-attention layers. Each MoE layer is followed by one sequential memory layer, and a self-attention layer is inserted after each 10 consecutive MoE and sequential memory layers. In our experiments, we vary the number of experts of MoE layers to be 2, 4 and 8, which are marked as MoE-2e, MoE-4e and MoE-8e respectively. The shared embedding network is a static model without MoE layers but a similar structure to the backbone.
%In this paper, MoE layers with 1, 2, 4 and 8 experts are used. We call the resulting models MoE-1e, MoE-2e, MoE-4e and MoE-8e. The shared embedding network is similar with the MoE-1e while have no router layer.

%\subsubsection{Acoustic model of baselines}

In our study, we built two baseline systems for evaluating the performance of our proposed method:

\begin{itemize}

\item[-] Baseline 1 (B1): The static model without MoE layers but a similar structure to the backbone of SpeechMoE models, which can also be treated as MoE-1e. Since the proposed method uses an extra embedding network, B1 model is designed to have 60 layers to be FLOP-matched with our MoE models.

\item[-] Baseline 2 (B2): The model with 4 experts in each MoE layer, which does not have the shared embedding network and is trained with only the auxiliary balancing loss proposed in Switch Transformer.

\end{itemize}
For all experiments on MoE models, we set the hyper-parameters $\alpha=0.1$, $\beta=0.1$ and $\gamma=0.01$.
%%We use CTC models for acoustic modeling.
%To illustrate the efficacy of our approach, 
%For the first experiment, we present our work with DFSMN-SAN with persistent memory model \cite{you2020dfsmn}. The DFSMN system uses 30 DFSMN components of 1024 hidden units, each with a projection layer of 512 units. 
%The look-back order and look ahead order of each memory block is 5 and 1 respectively, and the strides are 2 and 1 respectively. 
%The self-attention model contains 10 multi-head self-attention sublayers with a comparable size with 30 DFSMN model. We set the model dimension $d = 512$ and the number of heads $h = 8$. The DFSMN-SAN model consists of 30 DFSMN components and 3 multi-head self-attention sublayers. 

%For the second experiment, we improve the DFSMN-SAN by augmenting the self-attention sublayer with persistent memory. We set the number of heads to 8 for key-value memory vectors. The position embedding is shared across all the heads.

\section{Experimental Results}
\subsection{Adding sparsity $L1$ loss}
In this section, we investigate the performance of adding the sparsity $L1$ loss in training. We have trained two baseline systems for this evaluation. The first baseline system(B1) is the static model trained based on $L_r$ loss and The other one(B2) is trained based on $L_r$ and $L_b$ loss mentioned above. Our result of adding sparsity $L1$ loss relative to B2 is marked as MoE-$L1$.

% In this section, we fist evaluate the performance of adding the sparseness loss for the training process. To conduct the evaluation, we create two baseline systems. The first baseline system is the B1 model based on $L_r$ loss. We train another baseline system (B2 model) based on $L_r$ and $L_b$ loss. 

\begin{table}
%\begin{table}[h]\small
\caption{\textit{Results of adding sparseness $L1$ loss. } }
\vspace{-10pt}
\label{tab:1}
\begin{center}
\scalebox{0.85}{
    \begin{tabular}{|c|c|c|c|c|c|c|}
    \hline
    \multirow{2}*{\small{Model}} & \multirow{2}*{\small{Params}} & \multirow{2}*{\small{FLOPs}} & \multicolumn{4}{c|}{\small{Test set}} \\
    \cline{4-7}
    & & &\small{Read} & \small{Chat} & \small{Spon} &  \small{AISHELL}  \\
    \hline
    \small{B1}    & 71M & 2.3B     & 2.0 & 22.92 & 24.95  & 4.52 \\
    
    \small{B2}      & 134M & 2.3B    & 1.81  & 22.49 & 24.90 & 4.50 \\
    
    \small{MoE-{$L1$}}  & 134M  & 2.3B & \bf{1.69}  & \bf{22.47} & \bf{24.70} & \bf{4.25}\\
    
     \hline
    \end{tabular}
}
\end{center}
\vspace{-5pt}
\end{table}

As shown in table 1, B2 performs a little better than B1 with more parameters and comparable computation cost. It is as expected that the MoE-$L1$ which uses both balancing loss and sparsity $L1$ loss achieves the best performance compared with two baseline systems. This indicates that the additional sparsity $L1$ loss brings about more sparsity to router probability distribution. The routers become more distinctive and specialized for varying input frames so that the model get a better performance. %Notably, the MoE-$L1$ model achieves up to 15.5\% relative CER improvement over the B1 system on the Read test set.

   %In table 1, line 2 presents the results of the baseline system (B1).  Line 3 presents the results of the second baseline system (B2). The last line presents the results of adding the sparseness $L1$ loss (MoE-$L1$). The results show that B2 performs better than B1 with comparable computational cost. As expected, the MoE-$L1$ which combines the balancing loss with sparseness $L1$ loss performs best compared with other models. This indicates that adding the sparseness $L1$ loss encourages to learn a few high probability of router activations for some part of the training sample and low probability of router activations for the rest of the training sample, which results in learning a sparse router activation. Notably, the MoE-$L1$ model achieves up to 15.5\% relative CER improvement over the B1 system on the Read test set.
   %This indicates that the relative dependency learned by self attention layers can improve the system performance while self-attention mechanism is not necessary for low-level front layers, and only adding a few self-attention layers to high level can achieve competitive performance. Notably, the DFSMN-SAN model achieves up to 34.4\% relative CER improvement over the baseline model on the Read test set.

\subsection{Augmenting shared embedding network}
In this section, we evaluate the performance of the new router architecture which concatenates the shared embedding with output of the previous layer as the input of the router. As can be observed in table 2, the proposed router architecture achieves lower character error rate comparing with MoE-$L1$ model. 

It is worthy to note that only using output of previous layer as input does not work very well, which contradict with the method used in \cite{fedus2021switch}. A reasonable explanation is that for language modeling, the word input as high-level representation already has good distinction,  while for speech recognition the spectrum input is low-level feature which can not provide enough distinction information for routers, so the shared embedding network which converts low-level features to high-level embedding, is necessary to help router attain better selecting effect.

%Intuitively, training samples of same sequence label have similar embeddings which encourages the router have similar expert decisions and vice versa.
\begin{table}
%\begin{table}[h]\small
\caption{\textit{Results of augmenting shared embedding network and utilizing mean importance loss. } }
%\vspace{1pt}
\vspace{-10pt}
\label{tab:1}
\begin{center}
\scalebox{0.85}{
\begin{tabular}{|c|c|c|c|c|c|c|}
\hline
\multirow{2}*{\small{Model}} & \multirow{2}*{\small{Params}} & \multirow{2}*{\small{FLOPs}} & \multicolumn{4}{c|}{\small{Test set}} \\
\cline{4-7}
&  & & \small{Read} & \small{Chat} & \small{Spon} &  \small{AISHELL}  \\
\hline
\small{MoE-{$L1$}}      & 134M    & 2.3B & 1.69  & 22.47 & 24.70 & 4.25 \\
\small{+{emb}}  & 170M  & 2.3B & \bf{1.63}  & \bf{22.15} & \bf{24.15} & \bf{4.16}\\
\small{+{imp loss}}  & 170M & 2.3B & \bf{1.58}  & \bf{21.57} & \bf{23.31} & \bf{4.00}\\
 \hline
\end{tabular}
}
\end{center}
\vspace{-5pt}
\end{table}

\subsection{Utilizing mean importance loss}
The last line of table 2 presents the effects of the mean importance loss in place of the balancing loss. We observe that the proposed loss can further achieve lower character error rate than MoE-$L1$ model with embedding network on the four test sets. Since the mean importance loss encourages all experts to have equal importance, it will help the routers dispatch input frames to experts in a balanced way, avoiding the situation that some experts get no samples for training. Thus, the experts will be more diverse and result in a better performance.

%\begin{table}
%\begin{table}[h]\small
%\caption{\textit{Results of adding mean importance loss. } }
%\vspace{1pt}
%\label{tab:1}
%\begin{center}
%\begin{tabular}{|c|c|c|c|c|c|}
%\hline
%\multirow{2}*{\small{Model}} & %\multirow{2}*{\small{Para}} & \multicolumn{4}{c|}{\small{Test set}} \\
%\cline{3-6}&  & \small{Read} & \small{Chat} & \small{Spon} &  \small{AISHELL}  \\
%\hline

%\small{MoE-{L1}-{emb}}  & 170M  & {1.63}  & {22.15} & {24.15} & {4.16}\\

%\small{+{imp loss}}  & 170M  & \bf{1.58}  & \bf{21.57} & \bf{23.31} & \bf{4.00}\\

 %\hline
%\end{tabular}
%\end{center}
%\vspace{-15pt}
%\end{table}

%\subsection{Joint training}
%To evaluate the performance of the deep switch model with CE loss. 

\subsection{Increasing the number of experts}
%In this section, we evaluate the performance of the deep switch model with CE loss. To conduct the evaluation, we increase the number of experts in the deep switch networks from 2 to 8. Table 1 shows the performance comparison of 3 different numbers of expert. We observe that Deep switch model can achieve lower dev loss than the baseline model. Specially, through our evaluations we find that big-deepMOE has lower dev loss that the small-deepMOE.

\begin{figure}[!tb]
\begin{minipage}[b]{1.0\linewidth}
  \centering
  \centerline{\includegraphics[width=7.5cm]{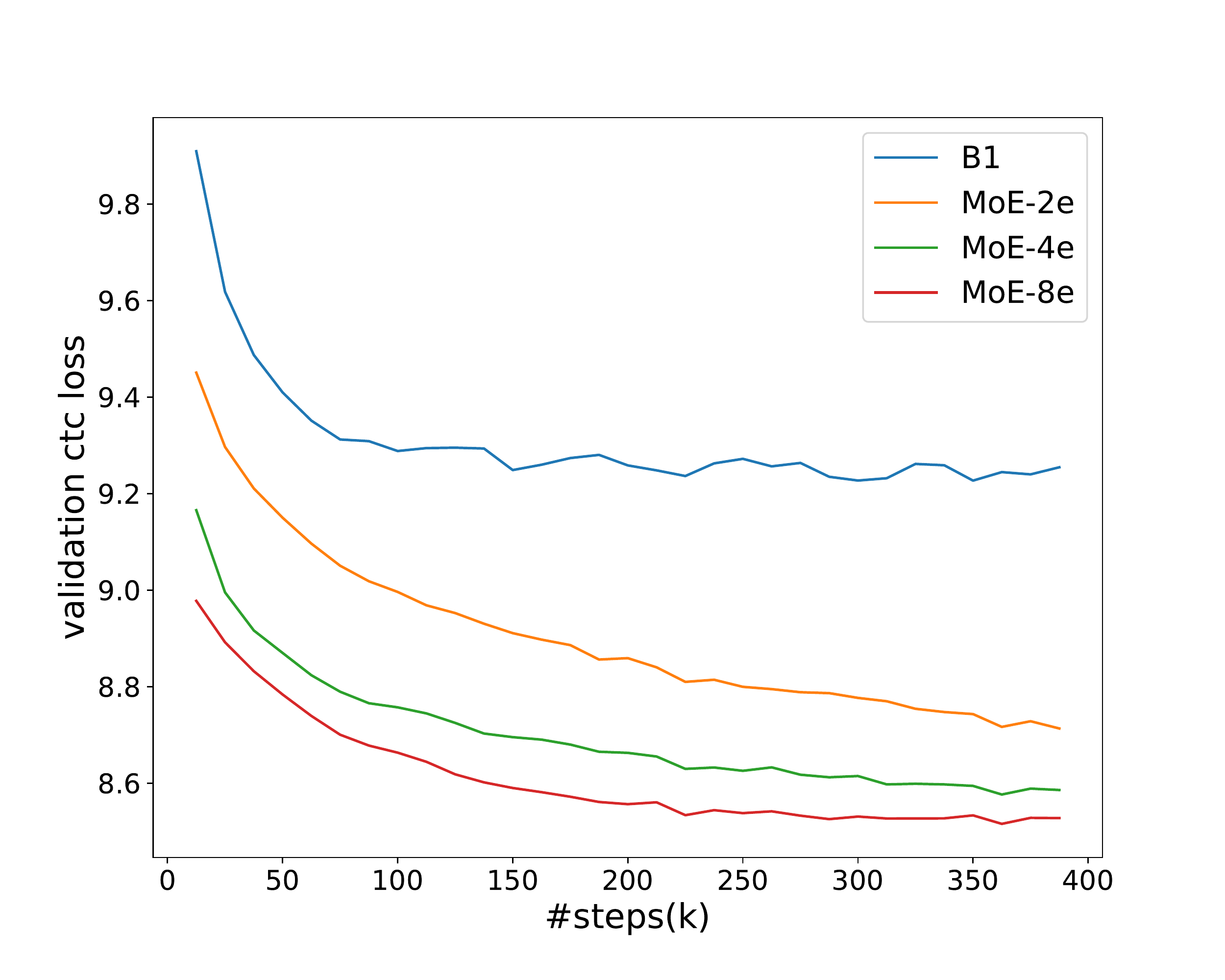}}
%  \vspace{2.0cm}
%  \centerline{(a) Result 1}\medskip
\end{minipage}
\vspace{-10pt}

\caption{Validation CTC loss for increasing expert number}
%\caption{Frames accuracy on development set against number of epochs for SGD, HF and SHF. ($N$T means SHF or HF is initialized with the result of SGD after $N$ epochs.)}
%\label{fig:res}
\end{figure}

In this section, we investigate the effect of increasing the number of experts. 
Table 3 shows the performance comparison on different number of experts with SpeechMoE. Line 2 presents the results of the baseline system (B1). The following three lines present results of 3 different number of experts which are marked as MoE-2e, MoE-4e and MoE-8e respectively. The results clearly show that performance get better as the number of experts increases. Specifically, MoE-8e achieves up to 23.0\% relative CER improvement over the baseline model on the Read test set, and the gain is between 7.0\%$ \sim$11.9\% for other more realistic test sets.

Figure 2 shows the validation CTC loss of MoE with different number of experts and the baseline model. As shown, the MoE-8e model produces the lowest CTC loss compared with both the baseline model and the other SpeechMoE models. Moreover, we observe that having more experts speeds up training. This suggests that increasing the number of expert leads to more powerful models.

\newcommand{\tabincell}[2]{\begin{tabular}{@{}#1@{}}#2\end{tabular}} 
\begin{table}
%\begin{table}[h]\small
\caption{\textit{Results of increasing the number of experts. } }
%\vspace{1pt}
\vspace{-10pt}
\label{tab:1}
\begin{center}
\scalebox{0.75}{
\begin{tabular}{|c|c|c|c|c|c|c|}
\hline
\multirow{2}*{\small{Model}} & \multirow{2}*{\small{Params}} & \multirow{2}*{\small{FLOPs}} & \multicolumn{4}{c|}{\small{Test set}} \\
\cline{4-7}
&  &  & \small{Read} & \small{Chat} & \small{Spon} &  \small{AISHELL}  \\
\hline
\small{B1}    & 71M     & 2.3B & 2.0 & 22.92 & 24.95  & 4.52 \\
\small{MoE-2e}    & 105M    & 2.3B & 1.62 & 21.82 & 23.52  & 4.08 \\

\small{MoE-4e}      & 170M  & 2.3B  & 1.58  & 21.57 & 23.31 & 4.00 \\

\small{MoE-8e} & {297M}  & 2.3B & \tabincell{c} {\bf{1.54} \\ \bf{(-23.0\%)}}  &  \tabincell{c} {\bf{21.31} \\ \bf{(-7.0\%)}} &  \tabincell{c} {\bf{22.97} \\  \bf{(-7.9\%)}} & \tabincell{c} {\bf{3.98} \\   \bf{(-11.9\%)}} \\

 \hline
\end{tabular}
}
\end{center}
\vspace{-15pt}
\end{table}
%\vspace{-15pt}

%Aliquam quis orci consectetur nulla luctus ullamcorper. Suspendisse finibus luctus erat a dapibus.

\section{Conclusions and future work}
In this paper, we explore a mixture of experts approach for speech recognition. 
We propose a novel dynamic routing acoustic model architecture, the router module is enhanced by combining the previous layer's output and embedding from an isolated embedding network. We also improve the training loss that can both achieve better sparsity and balancing among different experts. Thorough experiments are conducted on training with different loss and varied number of experts. Future work includes both extending training data scale and number of experts, increasing by one or two orders of magnitudes, and exploring the proposed SpeechMoE model with other end-to-end training framework such as transformer transducers.
\bibliographystyle{IEEEtran}

\bibliography{mybib}

\begin{thebibliography}{10}
\providecommand{\url}[1]{#1}
\csname url@samestyle\endcsname
\providecommand{\newblock}{\relax}
\providecommand{\bibinfo}[2]{#2}
\providecommand{\BIBentrySTDinterwordspacing}{\spaceskip=0pt\relax}
\providecommand{\BIBentryALTinterwordstretchfactor}{4}
\providecommand{\BIBentryALTinterwordspacing}{\spaceskip=\fontdimen2\font plus
\BIBentryALTinterwordstretchfactor\fontdimen3\font minus
  \fontdimen4\font\relax}
\providecommand{\BIBforeignlanguage}[2]{{%
\expandafter\ifx\csname l@#1\endcsname\relax
\typeout{** WARNING: IEEEtran.bst: No hyphenation pattern has been}%
\typeout{** loaded for the language `#1'. Using the pattern for}%
\typeout{** the default language instead.}%
\else
\language=\csname l@#1\endcsname
\fi
#2}}
\providecommand{\BIBdecl}{\relax}
\BIBdecl

\bibitem{DE2}
G.~E. Dahl, D.~Yu, L.~Deng, and A.~Acero, ``Context-dependent pre-trained deep
  neural networks for large-vocabulary speech recognition,'' in \emph{IEEE
  Transactions on audio, speech, and language processing}, vol.~20.\hskip 1em
  plus 0.5em minus 0.4em\relax IEEE, 2012, p. 30–42.

\bibitem{DE3}
D.~Yu and J.~Li, ``Recent progresses in deep learning based acoustic models,''
  in \emph{IEEE/CAA Journal of Automatica Sinica}, vol.~4.\hskip 1em plus 0.5em
  minus 0.4em\relax IEEE, 2017, p. 396–409.

\bibitem{sainath2013deep}
T.~N. Sainath, A.-r. Mohamed, B.~Kingsbury, and B.~Ramabhadran, ``Deep
  convolutional neural networks for lvcsr,'' in \emph{2013 IEEE international
  conference on acoustics, speech and signal processing}.\hskip 1em plus 0.5em
  minus 0.4em\relax IEEE, 2013, pp. 8614--8618.

\bibitem{CNN}
Y.~{Qian} and P.~C. {Woodland}, ``Very deep convolutional neural networks for
  robust speech recognition,'' in \emph{2016 IEEE Spoken Language Technology
  Workshop (SLT)}, 2016, pp. 481--488.

\bibitem{graves2013hybrid}
A.~Graves, N.~Jaitly, and A.-r. Mohamed, ``Hybrid speech recognition with deep
  bidirectional lstm,'' in \emph{2013 IEEE workshop on automatic speech
  recognition and understanding}.\hskip 1em plus 0.5em minus 0.4em\relax IEEE,
  2013, pp. 273--278.

\bibitem{ravanelli2018light}
M.~Ravanelli, P.~Brakel, M.~Omologo, and Y.~Bengio, ``Light gated recurrent
  units for speech recognition,'' \emph{IEEE Transactions on Emerging Topics in
  Computational Intelligence}, vol.~2, no.~2, pp. 92--102, 2018.

\bibitem{peddinti2015time}
V.~Peddinti, D.~Povey, and S.~Khudanpur, ``A time delay neural network
  architecture for efficient modeling of long temporal contexts,'' in
  \emph{Sixteenth Annual Conference of the International Speech Communication
  Association}, 2015.

\bibitem{zhang2018deep}
S.~Zhang, M.~Lei, Z.~Yan, and L.~Dai, ``Deep-fsmn for large vocabulary
  continuous speech recognition,'' in \emph{2018 IEEE International Conference
  on Acoustics, Speech and Signal Processing (ICASSP)}.\hskip 1em plus 0.5em
  minus 0.4em\relax IEEE, 2018, pp. 5869--5873.

\bibitem{transformer}
L.~{Dong}, S.~{Xu}, and B.~{Xu}, ``Speech-transformer: A no-recurrence
  sequence-to-sequence model for speech recognition,'' in \emph{2018 IEEE
  International Conference on Acoustics, Speech and Signal Processing
  (ICASSP)}, 2018, pp. 5884--5888.

\bibitem{emformer}
Y.~{Shi}, Y.~{Wang}, C.~{Wu}, C.-F. {Yeh}, J.~{Chan}, F.~{Zhang}, D.~{Le}, and
  M.~{Seltzer}, ``{Emformer: Efficient Memory Transformer Based Acoustic Model
  For Low Latency Streaming Speech Recognition},'' \emph{arXiv e-prints}, p.
  arXiv:2010.10759, Oct. 2020.

\bibitem{conformer}
A.~Gulati, J.~Qin, C.-C. Chiu, N.~Parmar, Y.~Zhang, J.~Yu, W.~Han, S.~Wang,
  Z.~Zhang, Y.~Wu, and R.~Pang, ``Conformer: Convolution-augmented transformer
  for speech recognition,'' 2020.

\bibitem{shoeybi2019megatron}
M.~Shoeybi, M.~Patwary, R.~Puri, P.~LeGresley, J.~Casper, and B.~Catanzaro,
  ``Megatron-lm: Training multi-billion parameter language models using model
  parallelism,'' \emph{arXiv preprint arXiv:1909.08053}, 2019.

\bibitem{brown2020language}
T.~B. Brown, B.~Mann, N.~Ryder, M.~Subbiah, J.~Kaplan, P.~Dhariwal,
  A.~Neelakantan, P.~Shyam, G.~Sastry, A.~Askell \emph{et~al.}, ``Language
  models are few-shot learners,'' \emph{arXiv preprint arXiv:2005.14165}, 2020.

\bibitem{jacobs1991adaptive}
R.~A. Jacobs, M.~I. Jordan, S.~J. Nowlan, and G.~E. Hinton, ``Adaptive mixtures
  of local experts,'' \emph{Neural computation}, vol.~3, no.~1, pp. 79--87,
  1991.

\bibitem{jordan1994hierarchical}
M.~I. Jordan and R.~A. Jacobs, ``Hierarchical mixtures of experts and the em
  algorithm,'' \emph{Neural computation}, vol.~6, no.~2, pp. 181--214, 1994.

\bibitem{lepikhin2020gshard}
D.~Lepikhin, H.~Lee, Y.~Xu, D.~Chen, O.~Firat, Y.~Huang, M.~Krikun, N.~Shazeer,
  and Z.~Chen, ``Gshard: Scaling giant models with conditional computation and
  automatic sharding,'' \emph{arXiv preprint arXiv:2006.16668}, 2020.

\bibitem{fedus2021switch}
W.~Fedus, B.~Zoph, and N.~Shazeer, ``Switch transformers: Scaling to trillion
  parameter models with simple and efficient sparsity,'' \emph{arXiv preprint
  arXiv:2101.03961}, 2021.

\bibitem{gross2017hard}
S.~Gross, M.~Ranzato, and A.~Szlam, ``Hard mixtures of experts for large scale
  weakly supervised vision,'' in \emph{Proceedings of the IEEE Conference on
  Computer Vision and Pattern Recognition}, 2017, pp. 6865--6873.

\bibitem{ahmed2016network}
K.~Ahmed, M.~H. Baig, and L.~Torresani, ``Network of experts for large-scale
  image categorization,'' in \emph{European Conference on Computer
  Vision}.\hskip 1em plus 0.5em minus 0.4em\relax Springer, 2016, pp. 516--532.

\bibitem{wang2020deep}
X.~Wang, F.~Yu, L.~Dunlap, Y.-A. Ma, R.~Wang, A.~Mirhoseini, T.~Darrell, and
  J.~E. Gonzalez, ``Deep mixture of experts via shallow embedding,'' in
  \emph{Uncertainty in Artificial Intelligence}.\hskip 1em plus 0.5em minus
  0.4em\relax PMLR, 2020, pp. 552--562.

\bibitem{cai2021dynamic}
S.~Cai, Y.~Shu, and W.~Wang, ``Dynamic routing networks,'' in \emph{Proceedings
  of the IEEE/CVF Winter Conference on Applications of Computer Vision}, 2021,
  pp. 3588--3597.

\bibitem{shazeer2017outrageously}
N.~Shazeer, A.~Mirhoseini, K.~Maziarz, A.~Davis, Q.~Le, G.~Hinton, and J.~Dean,
  ``Outrageously large neural networks: The sparsely-gated mixture-of-experts
  layer,'' \emph{arXiv preprint arXiv:1701.06538}, 2017.

\bibitem{graves2006connectionist}
A.~Graves, S.~Fern{\'a}ndez, F.~Gomez, and J.~Schmidhuber, ``Connectionist
  temporal classification: labelling unsegmented sequence data with recurrent
  neural networks,'' in \emph{Proceedings of the 23rd international conference
  on Machine learning}, 2006, pp. 369--376.

\bibitem{syllable}
Z.~Qu, P.~Haghani, E.~Weinstein, and P.~Moreno, ``Syllable-based acoustic
  modeling with ctc-smbr-lstm,'' in \emph{Automatic Speech Recognition and
  Understanding Workshop (ASRU), 2017 IEEE}.\hskip 1em plus 0.5em minus
  0.4em\relax IEEE, 2017, pp. 173--177.

\bibitem{far2}
I.~Himawan, P.~Motlicek, D.~Imseng, B.~Potard, N.~Kim, and J.~Lee, ``Learning
  feature mapping using deep neural network bottleneck features for distant
  large vocabulary speech recognition,'' in \emph{International Conference on
  Acoustics, Speech and Signal Processing}, 2015.

\bibitem{fsmn}
S.~{Zhang}, M.~{Lei}, Z.~{Yan}, and L.~{Dai}, ``Deep-fsmn for large vocabulary
  continuous speech recognition,'' in \emph{2018 IEEE International Conference
  on Acoustics, Speech and Signal Processing (ICASSP)}, 2018, pp. 5869--5873.

\bibitem{you2020dfsmn}
Z.~You, D.~Su, J.~Chen, C.~Weng, and D.~Yu, ``Dfsmn-san with persistent memory
  model for automatic speech recognition,'' in \emph{ICASSP 2020-2020 IEEE
  International Conference on Acoustics, Speech and Signal Processing
  (ICASSP)}.\hskip 1em plus 0.5em minus 0.4em\relax IEEE, 2020, pp. 7704--7708.

\end{thebibliography}

% \begin{thebibliography}{9}
% \bibitem[1]{Davis80-COP}
%   S.\ B.\ Davis and P.\ Mermelstein,
%   ``Comparison of parametric representation for monosyllabic word recognition in continuously spoken sentences,''
%   \textit{IEEE Transactions on Acoustics, Speech and Signal Processing}, vol.~28, no.~4, pp.~357--366, 1980.
% \bibitem[2]{Rabiner89-ATO}
%   L.\ R.\ Rabiner,
%   ``A tutorial on hidden Markov models and selected applications in speech recognition,''
%   \textit{Proceedings of the IEEE}, vol.~77, no.~2, pp.~257-286, 1989.
% \bibitem[3]{Hastie09-TEO}
%   T.\ Hastie, R.\ Tibshirani, and J.\ Friedman,
%   \textit{The Elements of Statistical Learning -- Data Mining, Inference, and Prediction}.
%   New York: Springer, 2009.
% \bibitem[4]{YourName17-XXX}
%   F.\ Lastname1, F.\ Lastname2, and F.\ Lastname3,
%   ``Title of your INTERSPEECH 2021 publication,''
%   in \textit{Interspeech 2021 -- 20\textsuperscript{th} Annual Conference of the International Speech Communication Association, September 15-19, Graz, Austria, Proceedings, Proceedings}, 2020, pp.~100--104.
% \end{thebibliography}

\end{document}